\AddToHook{package/hyperref/before}{\RequirePackage{float}}
\documentclass[submission]{eptcs}
\usepackage{amsmath}
\usepackage{listings}
\usepackage{graphicx}
\usepackage{hyperref}
\usepackage{underscore} 
\usepackage{cite}

\usepackage{amssymb}
\usepackage{xcolor}
\usepackage{tikz}
\usetikzlibrary{positioning,arrows.meta}

\lstdefinelanguage{Promela}{
  morekeywords={proctype,active,init,typedef,mtype,chan,of,do,od,if,fi,
    goto,break,unless,atomic,d_step,inline,run,skip,assert,printf,
    byte,bool,bit,short,int,unsigned,pid,ltl,hidden,show,local,
    len,empty,nempty,full,nfull,eval,timeout,true,false},
  sensitive=true,
  morecomment=[l]{//},
  morecomment=[s]{/*}{*/},
  morestring=[b]",
}
\definecolor{promkw}{RGB}{0,0,180}
\definecolor{promcomment}{RGB}{110,110,110}
\definecolor{promstr}{RGB}{0,120,110}
\title{Verifying Graceful Degradation in a Distributed Malware-Detection System with SPIN}
\author{Andrei ALDEA
\institute{Bitdefender, Ia\c{s}i, Romania\\
Alexandru Ioan Cuza University, Ia\c{s}i, Romania}
\email{aaldea@bitdefender.com}
\and
Dumitru-Bogdan PRELIPCEAN
\institute{
Bitdefender, Ia\c{s}i, Romania\\
Alexandru Ioan Cuza University, Ia\c{s}i, Romania}
\email{bprelipcean@bitdefender.com}}
\def\authorrunning{A. Aldea and D.B. Prelipcean}
\def\titlerunning{Verifying Graceful Degradation in Malware Detection with SPIN}
\def\event{FROM 2026}

\begin{document}

\maketitle

\begin{abstract}
Modern endpoint malware detection is distributed: a lightweight agent on each endpoint collects features from a scanned file or process, sends them to a remote server for analysis, and then \emph{enforces} a malicious verdict locally by blocking, quarantining, or disinfecting. Because the endpoint acts on the verdict, the distributed machinery surrounding detection must never turn a transient server failure into a wrong action. We present a formal model, in Promela, of the endpoint decision pipeline of such a system, abstracted from a production architecture at Bitdefender. The model captures the system's graceful-degradation fallback chain: when the primary analysis server times out, the endpoint falls back to an older legacy-protocol server, and failing that to a reduced-signature local scan, before committing a terminal verdict. Under explicit correctness assumptions for the detection engines, we specify six safety and liveness properties in linear temporal logic (LTL) and verify them exhaustively with the SPIN model checker. We show that the fallback machinery never causes a \emph{false positive} (an enforcement action against a benign file), never performs duplicate enforcement for one scan, weakens detection strength only in an explicit and ordered way, and always reaches a terminal verdict. A separate invalid-end-state search establishes deadlock freedom. We also check that the relevant behaviors are reachable and report how the state space grows with concurrent scans and endpoints. The work shows how model checking can give strong correctness guarantees for the failure-handling logic of a production security system, a layer that has received little direct formal attention.
\end{abstract}

\section{Introduction}
Endpoint malware detection safeguards digital infrastructures against a continually evolving landscape of cyber threats. To meet the twin demands of deep analysis and low endpoint overhead, modern detection architectures are distributed: a lightweight agent on each endpoint collects features from a scanned file or process, such as file metadata, process information, or behavioral telemetry, and offloads the computationally intensive analysis to a remote server. The server applies signature matching, heuristics, or machine-learning classifiers to these features and returns a verdict. Because feature extraction is cheap and analysis is expensive, this split keeps the endpoint light while centralizing detection capability.

Crucially, the verdict is not merely informational. On a malicious verdict the endpoint takes a \emph{local enforcement action}, dictated by policy: it blocks the offending process, quarantines the file, or disinfects it. The endpoint therefore acts on what the server tells it, which turns the reliability of the distributed machinery into a correctness concern. If a transient server failure causes the endpoint to act on the wrong verdict, the consequences are concrete: a benign file blocked (a \emph{false positive}, breaking legitimate software) or a threat allowed to run (a \emph{false negative}).

Servers do fail. They experience overload, maintenance windows, and targeted attacks, and a request may time out. Production systems answer this with \emph{graceful degradation}: rather than hang or guess, the endpoint follows a fallback chain. In the architecture we study, abstracted from a production system at Bitdefender, a request that the primary server does not answer in time is retried against an older legacy-protocol server, and if that also times out the endpoint runs a reduced-signature scan locally before committing a verdict. This fallback chain is where correctness is subtle: a response associated with an abandoned attempt must not be acted upon; detection strength must degrade only in bounded, explicit ways; and the endpoint must always reach a decision without deadlocking. These are exactly the behaviors that empirical testing, limited to the scenarios it can enumerate, is poorly suited to certify, and that model checking, by exhaustively exploring all interleavings, can.

In this paper we build a formal model of this endpoint decision pipeline in Promela and verify it with the SPIN model checker. Our contributions are:
\begin{itemize}
    \item A Promela model of the endpoint decision pipeline of a distributed malware-detection system, abstracted from a production Bitdefender architecture, including the primary/legacy/local fallback chain, timeout-based failure handling with discarding of late responses, and the local enforcement action that makes a verdict consequential (Section~\ref{sec:model}).
    \item A precise formulation in LTL of what correctness means for this pipeline: six safety and liveness properties organized around the principle that the distributed machinery must never turn a server failure into a wrong enforcement action (Section~\ref{sec:results}).
    \item An exhaustive SPIN verification of the full property set on the core, streaming, and concurrent configurations, including reachability witnesses and a separate deadlock analysis, with an honest account of how the state space grows and where it stops being tractable.
    \item An articulation of detection-\emph{infrastructure} correctness, the failure-handling logic between a verdict and the action taken on it, as a verification target in its own right, distinct from verifying malware behavior or detection rules.
\end{itemize}

We deliberately verify the detection \emph{infrastructure}, not malware behavior and not a new detection algorithm: the object of study is the failure-handling logic that sits between a verdict and the action taken on it. As we show, model checking exposes subtle failure interleavings in this logic and certifies that, under the explicit detection contracts A1--A2, the pipeline is safe.
\section{Preliminaries}
\label{sec:prelim}
We verify the model with SPIN, the explicit-state model checker for Promela~\cite{holzmann2004spin}. This section fixes the fragment of Promela used in the listings and the temporal operators used in the properties, so that the paper is self-contained for readers unfamiliar with either.

\paragraph{Promela.}
A Promela model is a set of concurrent processes (\texttt{proctype}s) that communicate over message channels and shared variables. A channel is declared with a capacity, e.g. \texttt{chan c = [4] of \{byte, mtype\}}: a positive capacity gives an asynchronous, buffered FIFO channel (capacity zero would give synchronous rendezvous, which we do not use). A tuple is sent with $c!v_1,\ldots,v_n$ and received with $c?v_1,\ldots,v_n$. A send on a full channel and a receive on an empty channel are not \emph{executable}; a process that reaches such a statement \emph{blocks} until it becomes executable. Control flow uses guarded commands: in \texttt{if :: g1 -> s1 :: g2 -> s2 fi} and its looping form \texttt{do ... od}, an option is eligible only when its guard \texttt{g} is executable, and SPIN explores \emph{every} eligible option, which is how nondeterminism, and hence both scheduling and injected faults, is modeled. A guard may be a boolean condition (the process blocks until it holds) or a channel operation. We also use \texttt{mtype} for symbolic constants, \texttt{atomic\{...\}} to execute a block without interleaving, and the channel predicate \texttt{empty(c)}. SPIN searches the entire reachable state space and reports assertion violations, deadlocks (states with no executable transition, called \emph{invalid end states}), and violations of temporal properties.

The following small exchange illustrates the same request--response shape used by the Primary path. Each statement is annotated so that the later model listings can be read without prior Promela experience.
\begin{lstlisting}
chan req = [1] of { byte, mtype };   /* buffered (id, input) tuples */
chan rsp = [1] of { byte, mtype };   /* buffered (id, verdict) tuples */
proctype Endpoint() {
    byte rid = 1; mtype verdict;     /* local request id and reply slot */
    req!rid, BENIGN;                 /* send both fields in one message */
    rsp?eval(rid), verdict;          /* accept only the matching id      */
}
proctype Server() {
    byte rid; mtype input;
    req?rid, input;                  /* receive fields into variables    */
    rsp!rid, CLEAN;                  /* return a correlated verdict      */
}
\end{lstlisting}
Here \texttt{eval(rid)} means ``match the received field against the current value of \texttt{rid}'', rather than storing the incoming field in that variable.

\paragraph{Temporal properties.}
Correctness properties are stated in Linear Temporal Logic (LTL)~\cite{emersonLTL90} over the model's global variables. We use two operators: $\Box\,\varphi$ (``always'' $\varphi$, i.e. $\varphi$ holds in every reachable state) for \emph{safety} (``something bad never happens''), and $\Diamond\,\varphi$ (``eventually'' $\varphi$) for \emph{liveness} (``something good eventually happens''). Most of our properties have the form $\Box\,(a \rightarrow b)$, ``whenever $a$ holds, $b$ holds'', and one has the form $\Box\,(a \rightarrow \Diamond\, b)$, ``every $a$ is eventually followed by $b$''. SPIN checks such a property by searching for a counterexample execution; finding none certifies that the property holds over all executions of the model.
\section{Related Work}
Formal methods, and model checking in particular, have been applied widely to security-critical and distributed systems to verify safety, liveness, and fault-tolerance. We position our work along three lines: model checking of malware behavior, verification of distributed systems in general, and other formal treatments of malware detection.

\paragraph{Model checking and malware.}
A prominent line of work uses model checking to analyze the \emph{behavior of a binary}: Song and Touili model programs as pushdown systems and specify malicious behaviors in temporal logic, deciding whether a sample exhibits them~\cite{song2012efficient,song2014model}. This is a fundamentally different problem from ours. There, the object of verification is the malware; the question is whether a given program is malicious. Here, the object is the \emph{detection infrastructure}; the question is whether the distributed system that carries verdicts and acts on them stays correct under failure. The two are complementary: one certifies what should be detected, the other certifies that the machinery around detection behaves safely.

\paragraph{SPIN for distributed and security-critical systems.}
SPIN is widely used to analyze concurrent systems and expose subtle design flaws~\cite{holzmann2004spin}, from routing and telecom protocols~\cite{kaur-2012} to BPEL Web-service flows translated into Promela for verification~\cite{nakashiro2011translation}. Xiao et al. extend SPIN-based security analysis to composed protocols, using message-field detection and component recognition to alleviate state-space explosion~\cite{xiao2022composition}. Promela/SPIN has also modeled Modbus communication~\cite{nardone2016modbus} and Ethereum smart-contract interaction models verified in LTL~\cite{yang2022formal}. State-space explosion remains the central obstacle, and partial-order reduction is a standard mitigation~\cite{clarke2012stateexplosion}; our own scalability wall (Section~\ref{sec:results}) is an instance of the same phenomenon.

\paragraph{Formal verification of modern distributed systems.}
Formal methods are increasingly used on production distributed systems. Amazon Web Services applies model checking, property-based testing, and formal specification to services such as S3~\cite{aws-correctness}. Recent work spans TLA+ and automated testing for the Confidential Consortium Framework~\cite{HowardKACC25}, IVy proofs of consensus safety in decentralized finance~\cite{PraveenRD24}, formal methods for hardware security~\cite{abs-2505-11963}, language-agnostic certification of message-passing protocol compliance~\cite{zhang2025languageagnosticlogicalrelationmessagepassing}, and multi-grained specifications for ZooKeeper that trade specification granularity against scalability~\cite{arxiv-multigrained}. For unbounded system sizes, parameterized techniques such as extended threshold automata verify round-based algorithms for all $N$~\cite{BaumeisterEJSV24}; we return to these as the principled route past our finite-instance results.

\paragraph{Verifying detection rules.}
Within the same application domain but on a different problem, Prelipcean and Dima verify detection \emph{rules} rather than the infrastructure that runs them: they give rules a compositional operational semantics and check their conformance to threat models expressed as attack trees, using bisimulation and weak trace inclusion in the CADP toolbox~\cite{cadp}, on malware such as LokiBot and Emotet~\cite{prelipcean2025bridging}. That is an equivalence question about detection logic, for which process algebra is a natural fit; ours is a question about the asynchronous behavior of the surrounding system under failure, for which explicit-state model checking with SPIN and LTL is a natural fit. Beyond sharing the malware-detection domain, the two have little in common in content or technique. Our untimed treatment of deadlines (Section~\ref{sec:model}) is deliberate: none of our properties depend on real durations, so deadline expiration is abstracted as a nondeterministic abandonment transition. Quantitative timing would instead call for a timed-automata tool such as UPPAAL~\cite{uppaal}.

In summary, while model checking has been applied both to malware behavior and to distributed systems at large, the failure-handling \emph{infrastructure} of a distributed malware-detection system has received little direct attention. This paper addresses that gap.

\section{System Architecture}
\label{sec:arch}
The system we model is abstracted from a production endpoint-security architecture at Bitdefender; it is not a hypothetical example. We keep only the elements that bear on the correctness of the failure-handling logic, and elide the detection algorithms themselves. We first describe the modern one-shot detection protocol and the legacy protocol it evolved from, which survives in the system as a fallback, and then the graceful-degradation chain that ties them together with a local reduced scan.

\subsection{From ping-pong to one-shot analysis}
\begin{figure}[tbp]
    \centering
    \begin{tikzpicture}[>=Stealth, font=\small,
        head/.style={draw, rounded corners, minimum width=2.1cm, minimum height=0.7cm},
        msg/.style={->, thick},
        note/.style={draw, fill=black!5, rounded corners, font=\footnotesize, anchor=west},
        lbl/.style={font=\footnotesize, midway, above}]
        \node[head] (ep) at (0,0) {Endpoint};
        \node[head] (sv) at (8,0) {Server};
        \foreach \n in {ep,sv} { \draw[dashed] (\n.south) -- ++(0,-4.6); }
        \draw[msg] (0,-0.7) -- node[lbl]{whole file} (8,-0.7);
        \node[note] at (8.15,-1.15) {initial analysis};
        \draw[msg] (8,-1.8) -- node[lbl]{request feature} (0,-1.8);
        \draw[msg] (0,-2.55) -- node[lbl]{feature data} (8,-2.55);
        \node[font=\footnotesize] at (4,-3.0) {\ldots\ repeats per feature};
        \draw[msg] (8,-3.75) -- node[lbl]{verdict} (0,-3.75);
    \end{tikzpicture}
    \caption{The legacy ping-pong protocol: after the initial file upload, the server repeatedly requests additional features from the endpoint until it can decide, then returns a verdict.}
    \label{fig:traditional-sequence}
\end{figure}
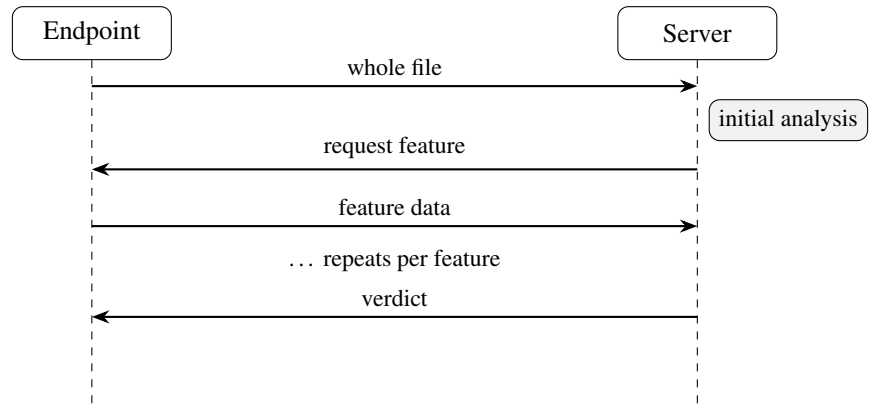

In the legacy protocol, illustrated in Figure~\ref{fig:traditional-sequence}, the endpoint first sends a file to the server, and the server then issues a sequence of follow-up requests for additional dynamic features (runtime behavior, process information) that it cannot infer from the static file. Each request is a separate round trip: the endpoint extracts the requested data and returns it, and this ping-pong continues until the server has enough information to decide. This design has three costs: repeated round trips inflate detection latency; re-transmitting data consumes bandwidth across many endpoints; and the ad-hoc, per-feature requests lack a unified serialization format, complicating client--server integration.

The modern protocol removes these costs by inverting the flow. The endpoint runs its local feature-collection stages to completion, then sends the complete feature set collected for that file to the server in a single request; the server replies once with a verdict. Feature extraction is cheap and local; deep analysis is expensive and remote; sending everything once, rather than negotiating feature by feature, is what makes the split efficient. The Legacy protocol need not receive exactly the same evidence: it starts from its initial request and obtains additional, potentially different features through server-initiated RPCs until it can decide. Thus, the two protocols differ both in message structure and in the evidence delivered on a particular execution.

\subsection{Actors}
The abstracted system has one active agent and three detection oracles of decreasing capability:
\begin{itemize}
    \item \textbf{Endpoint.} Runs on each host. It collects the feature set locally and, on a malicious verdict, performs the policy-dictated \emph{enforcement action}: blocking the process, quarantining the file, or disinfecting it.
    \item \textbf{Primary Server.} Answers one-shot requests using the full signature and heuristic database.
    \item \textbf{Legacy Server.} The older server, still deployed, speaking the ping-pong protocol and requesting additional features through RPC calls before returning a verdict.
    \item \textbf{Reduced local scan.} An on-endpoint scan against a reduced signature database. It runs without any server and is strictly weaker: it may miss malware the servers would catch.
\end{itemize}

\subsection{Graceful degradation}
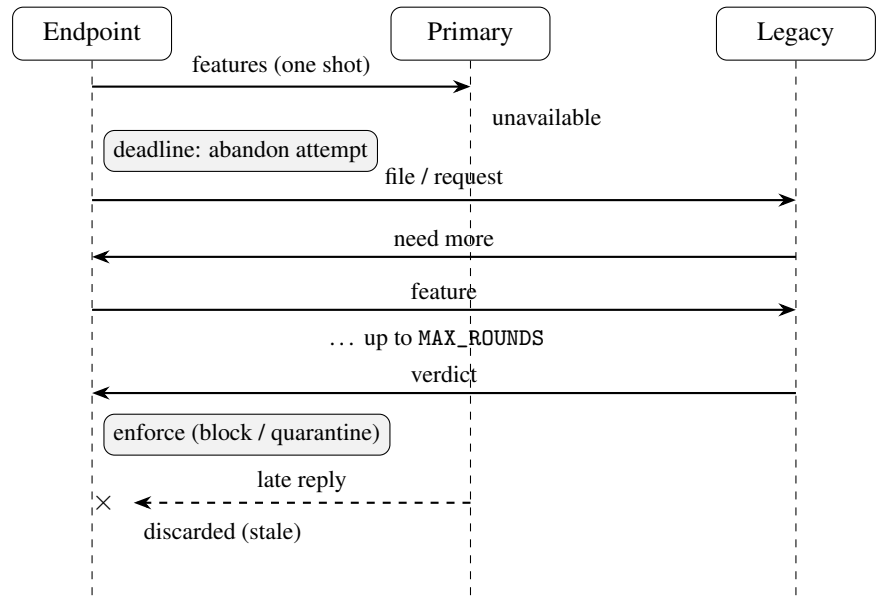
\begin{figure}[htbp]
    \centering
    \begin{tikzpicture}[>=Stealth, font=\small,
        head/.style={draw, rounded corners, minimum width=2.1cm, minimum height=0.7cm},
        msg/.style={->, thick},
        note/.style={draw, fill=black!5, rounded corners, font=\footnotesize, anchor=west},
        lbl/.style={font=\footnotesize, midway, above}]
        \node[head] (ep) at (0,0)   {Endpoint};
        \node[head] (pr) at (5,0)   {Primary};
        \node[head] (lg) at (9.3,0) {Legacy};
        \foreach \n in {ep,pr,lg} { \draw[dashed] (\n.south) -- ++(0,-7.1); }
        \draw[msg] (0,-0.7) -- node[lbl]{features (one shot)} (5,-0.7);
        \node[font=\footnotesize, anchor=west] at (5.15,-1.1) {unavailable};
        \node[note] at (0.15,-1.55) {deadline: abandon attempt};
        \draw[msg] (0,-2.2) -- node[lbl]{file / request} (9.3,-2.2);
        \draw[msg] (9.3,-2.95) -- node[lbl]{need more} (0,-2.95);
        \draw[msg] (0,-3.65) -- node[lbl]{feature} (9.3,-3.65);
        \node[font=\footnotesize, anchor=west] at (3.0,-4.05) {\ldots\ up to \texttt{MAX\_ROUNDS}};
        \draw[msg] (9.3,-4.75) -- node[lbl]{verdict} (0,-4.75);
        \node[note] at (0.15,-5.3) {enforce (block / quarantine)};
        \draw[msg, dashed] (5,-6.2) -- node[lbl]{late reply} (0.55,-6.2);
        \node[font=\normalsize] at (0.15,-6.2) {$\times$};
        \node[font=\footnotesize, anchor=west] at (0.55,-6.6) {discarded (stale)};
    \end{tikzpicture}
    \caption{The fallback chain under failure. The Endpoint sends the complete feature set to the Primary in one message; on deadline expiration it falls back to the Legacy, which drives a bounded ping-pong negotiation before returning a verdict. The Endpoint commits that verdict and enforces it only if malicious. A response visible after the abandoned Primary attempt is discarded as stale. The reduced local scan (a further, server-less fallback) is omitted for clarity.}
    \label{fig:fallback}
\end{figure}

Servers can be overloaded, under maintenance, or under attack, so a request may time out. Rather than hang or guess, the endpoint follows the fallback chain of Figure~\ref{fig:fallback}. It sends the feature set to the Primary Server; if no verdict arrives within the deadline, it tries the Legacy Server under one overall attempt deadline; if that too expires, it runs the Reduced local scan, which always yields a verdict. The Endpoint commits the first accepted terminal verdict and performs a local enforcement action exactly when that verdict is malicious.

The verification is conditional on two explicit detection contracts. \emph{A1 (full-server correctness)} states that, whenever Primary or Legacy responds, its verdict equals the file's independent ground truth; this assumption applies even though their feature evidence may differ. \emph{A2 (reduced-scan behavior)} states that Reduced-local never flags a benign file, but may either detect or miss a malicious file. We verify the distributed failure-handling machinery under A1--A2, rather than the detection algorithms themselves.

The transport is modeled as reliable: messages are neither lost, duplicated, nor retransmitted at the application level. Deadline expiration, the only communication failure exposed at this abstraction level, is represented without a clock. For safety we conservatively allow a response to become visible after its attempt has been abandoned; its attempt identifier makes it stale, so it is discarded rather than committed.

\section{Promela Model}
\label{sec:model}

We model the endpoint decision pipeline of Section~\ref{sec:arch} as three Promela processes: one \textbf{Endpoint}, a \textbf{PrimaryServer} (one-shot protocol), and a \textbf{LegacyServer} (ping-pong protocol). There is no load balancer and no server pool: the interesting structure is the heterogeneous fallback chain, not horizontal distribution. The Reduced local scan involves no communication and is an inline terminal step of the Endpoint. Processes communicate only over channels, never by writing each other's variables.

\subsection{State and the detection abstraction}
We abstract away feature extraction and the detection algorithms themselves, and keep only what the correctness properties depend on: the file's independent ground truth, which oracle answered, and the resulting verdict and enforcement. A scan's ground truth is a symbolic value chosen non-deterministically at start-up, so both a benign and a malicious file are covered by every verification run:

\begin{lstlisting}
mtype = { BENIGN, MALICIOUS, CLEAN, MAL, NEED_MORE,
          PRIMARY, LEGACY, REDUCED, NONE };
mtype ground_truth = NONE;     /* set non-deterministically in init */
bool  enforced = false;        /* an enforcement action was taken */
byte  enforce_count = 0;       /* number of enforcement actions   */
\end{lstlisting}

The complete observable state also records \texttt{committed\_verdict} (the terminal \texttt{CLEAN} or \texttt{MAL} verdict), \texttt{answered\_by} (Primary, Legacy, or Reduced-local), \texttt{primary\_timed\_out} and \texttt{legacy\_timed\_out} (whether each attempt was abandoned), \texttt{reached\_reduced} (whether the weakest mode was entered), and \texttt{terminated} (whether the scan reached its single commit point). The Boolean \texttt{enforced} and counter \texttt{enforce\_count} record whether, and how many times, a malicious verdict caused a local action.

The model instantiates A1 by mapping benign inputs to \texttt{CLEAN} and malicious inputs to \texttt{MAL} at both full servers. Under A2, Reduced-local also maps benign inputs to \texttt{CLEAN}, but maps a malicious input nondeterministically to \texttt{MAL} or \texttt{CLEAN}. These are assumptions about the detection oracles; the properties verify that the surrounding control flow preserves their outcomes.

\subsection{Servers and the fault model}
The Primary server answers a request in a single exchange, or stays silent. The silent branch is the fault model: a non-deterministic \texttt{skip} that produces no reply, capturing a server that is down or too slow for this request (and, per request, one that recovers on the next).

\begin{lstlisting}
proctype PrimaryServer() {
    byte aid; mtype gt;
endPrimary:
    do
    :: to_primary?aid, gt ->
        if
        :: gt == MALICIOUS -> to_endpoint!aid, MAL;   /* full DB */
        :: gt != MALICIOUS -> to_endpoint!aid, CLEAN; /* A1 */
        :: skip;                                      /* unavailable */
        fi;
    od;
}
\end{lstlisting}

The Legacy server speaks the ping-pong protocol: for each request it drives a bounded feature negotiation. In each round it may ask the Endpoint for another feature (\texttt{NEED\_MORE}) and wait for the reply, decide and return a verdict, or go silent. Production executions perform multiple RPCs, but their precise minimum is not part of the abstraction: the model conservatively permits between zero and \texttt{MAX\_ROUNDS} feature requests. The properties depend on bounded negotiation and correct message correlation, not on the minimum number of rounds.

\begin{lstlisting}
proctype LegacyServer() {
    byte aid; mtype gt; byte rounds;
endLegacy:
    do
    :: to_legacy?aid, gt ->               /* initial request */
        rounds = 0;
        do
        :: rounds < MAX_ROUNDS ->         /* ask for one more feature */
            to_endpoint!aid, NEED_MORE;
            if
            :: to_legacy_feat?eval(aid), gt -> rounds++;  /* feature arrived */
            :: true -> break;                             /* endpoint gone   */
            fi;
        :: gt == MALICIOUS -> to_endpoint!aid, MAL;   break;  /* decide */
        :: gt != MALICIOUS -> to_endpoint!aid, CLEAN; break;
        :: skip -> break;                 /* unavailable mid-exchange */
        od;
    od;
}
\end{lstlisting}

The Legacy channels represent distinct logical flows in the production RPC architecture. The client-initiated session request travels on \texttt{to\_legacy}; after that session is established, the Legacy initiates feature RPCs through \texttt{NEED\_MORE}, and the Endpoint's RPC replies travel on \texttt{to\_legacy\_feat}. They are separate Promela channels because they are different protocol operations, even if an implementation ultimately multiplexes them over lower-level transport machinery. The \texttt{eval(aid)} guard on the feature-reply receive accepts only a reply correlated with the active attempt.

Both the Legacy's wait for a feature and the Endpoint's wait for a server reply include an always-eligible abandonment branch. Without the former, the model could reach an invalid end state in which the Endpoint has timed out and completed its fallback chain while the Legacy remains blocked forever waiting for a feature reply that will never be sent. The \texttt{endPrimary}/\texttt{endLegacy} labels mark the servers' idle receive-loop heads as valid end states, rather than such abandoned mid-protocol waits.

\subsection{Endpoint: fallback with timeout and stale-reply discard}
The Endpoint tries the Primary with a one-shot wait that accepts the matching reply, discards stale replies from abandoned attempts, and may abandon the attempt nondeterministically:

\begin{lstlisting}
inline await_primary(aid, got, v) {
    got = false;
    do
    :: to_endpoint?aw_sid, aw_kind ->
        if
        :: aw_sid == aid -> v = aw_kind; got = true; break; /* verdict */
        :: else          -> skip;                           /* stale   */
        fi;
    :: true -> break;                                       /* deadline expires */
    od;
}
\end{lstlisting}

The always-enabled branch is an untimed abstraction of deadline expiration, not Promela's global \texttt{timeout} keyword and not a quantitative clock. It explores both relevant orderings: the reply is accepted first, or the attempt is abandoned first. The latter ordering conservatively permits a subsequently visible reply, which the identifier check discards. On abandonment the Endpoint falls back to the Legacy, where the wait is a loop: a \texttt{NEED\_MORE} is answered with a feature, a verdict ends the attempt, a stale reply is discarded, and expiration of the overall Legacy-attempt deadline abandons the attempt:

\begin{lstlisting}
attempt++; to_legacy!attempt, ground_truth;        /* initial request */
do
:: to_endpoint?aw_sid, aw_kind ->
    if
    :: aw_sid != attempt -> skip;                          /* stale: discard */
    :: aw_sid == attempt && aw_kind == NEED_MORE ->
        to_legacy_feat!attempt, ground_truth;              /* answer feature */
    :: aw_sid == attempt && aw_kind == MAL   -> v = MAL;   got = true; break;
    :: aw_sid == attempt && aw_kind == CLEAN -> v = CLEAN; got = true; break;
    fi;
:: true -> break;                                          /* deadline: abandon */
od;
\end{lstlisting}

If the Legacy attempt is also abandoned, the Endpoint runs the reduced scan, which always yields a verdict and may miss a malicious file under A2, then reaches the single commit point. A \texttt{MAL} verdict triggers enforcement; a \texttt{CLEAN} verdict does not:

\begin{lstlisting}
reached_reduced = true; answered_by = REDUCED;
if :: ground_truth == MALICIOUS ->
        if :: v = MAL :: v = CLEAN fi;             /* bounded, labeled miss */
   :: else -> v = CLEAN; fi;
commit:
    if :: v == MAL -> enforced = true; enforce_count++;   /* enforcement */
       :: else -> skip; fi;
\end{lstlisting}

Attempt identifiers increase monotonically across the Primary and Legacy attempts (and, in the streaming model, across scans), so a reply belonging to any earlier attempt fails the identifier check and cannot be mistaken for the current verdict. The enforcement action (block, quarantine, disinfect) is abstracted to setting \texttt{enforced}; what matters for correctness is only \emph{whether} and \emph{how often} it fires.

\subsection{Faithfulness of the abstraction}
Table~\ref{tab:faithful} maps each element of the real system to its Promela counterpart and states the abstraction made. The guiding principle is to model the failure-handling control flow precisely while abstracting detection content to symbolic values.

\begin{table}[ht!]
\centering
\caption{Mapping from the real system to the Promela model.}
\label{tab:faithful}
\begin{tabular}{|l|l|l|}
\hline
\textbf{Real element} & \textbf{Promela construct} & \textbf{Abstraction} \\
\hline
File ground truth       & \texttt{ground\_truth} (in \texttt{init}) & benign/malicious; both explored \\
Primary server          & \texttt{PrimaryServer}, one exchange & verdict equals ground truth (A1) \\
Legacy ping-pong        & \texttt{LegacyServer}, \texttt{NEED\_MORE} rounds & 0--\texttt{MAX\_ROUNDS}; minimum abstracted \\
Feature content         & symbolic reply on \texttt{to\_legacy\_feat} & content/sufficiency elided \\
Server down or slow     & non-deterministic \texttt{skip}   & crash/omission, per request \\
Request deadline        & \texttt{true -> break} while waiting & untimed abandonment \\
Late-reply handling     & identifier check \texttt{aw\_sid==aid} & stale verdicts discarded \\
Reduced local scan      & inline terminal branch           & no FP; may miss malware (A2) \\
Enforcement action      & \texttt{enforced}, \texttt{enforce\_count} & block/quarantine abstracted to a flag \\
\hline
\end{tabular}
\end{table}

\section{Results}
\label{sec:results}
We verify each property with SPIN over the model of Section~\ref{sec:model}. Because the ground truth of a scan is chosen non-deterministically, a single verification run covers both the benign and the malicious case; there is no need to enumerate inputs or generate per-input models. Each property is a separate LTL claim, checked in its own run. Safety properties are established by an exhaustive search for a violating state; the one liveness property is checked by an acceptance-cycle search.

\subsection{Properties}
We verify six properties, stated below over the model's global variables and classified by the specification pattern of Dwyer, Avrunin and Corbett~\cite{dwyer1999patterns}. They are organized around a single principle: the distributed machinery must never turn a server failure into a wrong enforcement action. In the streaming model each per-scan property is evaluated at a settled snapshot of the scan; we give the single-scan forms here.

\begin{description}
    \item[S1, No false positive (Absence).] An enforcement action is never taken against a benign file:
    \[ \Box\,(\mathit{enforced} \rightarrow \mathit{ground\_truth} = \mathrm{MALICIOUS}). \]
    Under A1--A2 this can only fail through the plumbing, so S1 certifies that the plumbing introduces no false positive.
    \item[S2, No duplicate enforcement (Universality).] At most one enforcement action is ever taken for a scan, even when a response becomes visible after its attempt was abandoned:
    \[ \Box\,(\mathit{enforce\_count} \le 1). \]
    \item[S3, Degradation order (Precedence).] Reduced-local is entered only after both server attempts have timed out:
    \[ \Box\,\big(\mathit{reached\_reduced} \rightarrow (\mathit{primary\_timed\_out} \wedge \mathit{legacy\_timed\_out})\big). \]
    \item[L1, Progress (Existence).] Every scan eventually reaches its terminal commit point:
    \[ \Diamond\, \mathit{terminated}. \]
    Because the commit point assigns one terminal verdict, L1 together with the Endpoint's single commit point establishes that every scan commits a verdict. This progress property is distinct from the invalid-end-state search for deadlock freedom reported below.
    \item[D1, Nominal soundness (Response).] A malicious file answered by a server is detected:
    \[
    \begin{aligned}
    \Box\,\big(&\mathit{terminated} \wedge \mathit{ground\_truth}{=}\mathrm{MALICIOUS} \\
    &{}\wedge \mathit{answered\_by} \in \{\mathrm{PRIMARY},\mathrm{LEGACY}\} \\
    &{}\rightarrow \mathit{committed}{=}\mathrm{MAL}\big).
    \end{aligned}
    \]
    \item[D2, Confined false negative (Precedence).] A malicious file escapes enforcement only if the scan fell through to the reduced local scan:
    \[
    \begin{aligned}
    \Box\,\big(& (\mathit{terminated} \wedge \mathit{ground\_truth}{=}\mathrm{MALICIOUS} \wedge \neg\mathit{enforced}) \\
    &{}\rightarrow \mathit{reached\_reduced}\big).
    \end{aligned}
    \]
\end{description}

Together, S1--S3 bound the damage a failure can do (no false positive, no duplicate enforcement, no premature weakening), L1 guarantees the pipeline always commits a terminal verdict, and D1--D2 confine every false negative to the single, explicitly weaker terminal mode.

\subsection{Verification results}
Table~\ref{tab:verification-results} reports the core configuration (one endpoint, one scan). All six properties hold, and each run completes in under a second.

\begin{table}[ht!]
\centering
\caption{SPIN verification of the six properties on the core configuration. All hold (0 errors).}
\label{tab:verification-results}
\begin{tabular}{|l|c|c|c|c|}
\hline
\textbf{Property} & \textbf{Class} & \textbf{States stored} & \textbf{Depth} & \textbf{Errors} \\
\hline
S1 no\_false\_positive   & safety   & 6{,}168 & 101 & 0 \\
S2 no\_duplicate\_enforcement & safety & 6{,}168 & 101 & 0 \\
S3 bounded\_order        & safety   & 6{,}168 & 101 & 0 \\
L1 progress              & liveness & 5{,}659 & 98 & 0 \\
D1 nominal\_soundness    & safety   & 6{,}168 & 101 & 0 \\
D2 confined\_fn          & safety   & 6{,}168 & 101 & 0 \\
\hline
\end{tabular}
\end{table}

A property of the form $\Box(a \rightarrow b)$ is vacuously true if $a$ is never reached, a common and easily overlooked failure of such specifications~\cite{kupferman1999vacuity}. We therefore check the implication antecedents and the representative paths needed to exercise the properties, using trap claims whose \emph{violation} exhibits the state of interest (Table~\ref{tab:vacuity}). Each trap is violated; the counterexample SPIN returns for a trap is a concrete witness (e.g.\ an execution in which enforcement occurs, or in which a malicious file is missed by the reduced scan).

\begin{table}[ht!]
\centering
\caption{Reachability checks. Each trap is violated, so a witness execution reaches the indicated behavior.}
\label{tab:vacuity}
\begin{tabular}{|l|l|}
\hline
\textbf{Trap claim (is violated)} & \textbf{Establishes reachability of} \\
\hline
$\Box\,\neg\,\mathit{enforced}$                                    & enforcement (S1, S2) \\
$\Box\,\neg\,\mathit{reached\_reduced}$                            & the reduced-scan fallback (S3, D2) \\
$\Box\,\neg(\mathit{enforced} \wedge \mathit{answered\_by}{=}\mathrm{LEGACY})$ & ping-pong server detection then enforcement (D1) \\
$\Box\,\neg(\mathit{terminated} \wedge \mathit{malicious} \wedge \neg\mathit{enforced})$ & a genuine false-negative run (D2) \\
\hline
\end{tabular}
\end{table}

Finally, separate claim-free searches with invalid-end-state detection enabled report no errors for any of the three configurations: no invalid end state is reachable. A server idling at its receive-loop head is a declared valid end state, so it is not mistaken for an abandoned mid-protocol wait.

\subsection{Scalability}
Beyond the core configuration (one endpoint, one scan), we verify two richer ones. The \emph{streaming} configuration gives the endpoint a stream of scans; since attempt identifiers stay unique across scans, a reply that arrives late during one scan is discarded during the next, a cross-scan race the single-scan model cannot exhibit. The \emph{concurrent} configuration adds a second, symmetric endpoint sharing the two servers. In both, per-scan observables are reset at each scan boundary and read only at the settled snapshot, so no property is evaluated straddling two scans. Verification runs cover all six properties on both the streaming and substantially larger concurrent configurations; each reports zero errors. Table~\ref{tab:performance-metrics} separates the representative safety search, the L1 acceptance-cycle search, and the claim-free invalid-end-state search. The growth comes from the added interleavings rather than from replicating identical components.

\begin{table}[ht!]
\centering
\caption{Verification across fixed configurations. Every LTL run and claim-free search reports 0 errors; cells show states stored/depth.}
\label{tab:performance-metrics}
\small
\begin{tabular}{|l|c|c|r|r|r|}
\hline
\textbf{Configuration} & \textbf{E$\times$S} & \textbf{LTL} &
\shortstack{\textbf{Safety}\\\textbf{stored/depth}} &
\shortstack{\textbf{L1}\\\textbf{stored/depth}} &
\shortstack{\textbf{Claim-free}\\\textbf{stored/depth}} \\
\hline
core       & $1\times1$ & 6/6 & 6{,}168/101 & 5{,}659/98 & 6{,}168/51 \\
streaming  & $1\times2$ & 6/6 & 314{,}527/209 & 305{,}176/206 & 314{,}527/105 \\
concurrent & $2\times1$ & 6/6 & 22{,}177{,}616/238 & 21{,}055{,}298/235 & 20{,}892{,}104/120 \\
\hline
\end{tabular}
\end{table}

Two remarks. First, the number of states is not a measure of a model's significance: faithful models of subtle concurrency properties are often compact, and published SPIN models of harder protocols verify in comparable or smaller spaces (a Byzantine reliable broadcast at $N{=}7$ in about $1.9\times10^{5}$ states~\cite{john2013fault}, Paxos instances in tens of states~\cite{delzanno2014paxos}). What matters is that the properties are non-vacuously exercised over all failure interleavings, which the trap checks confirm. Second, an attempted search with two endpoints each running two scans exhausted available memory before completion, so we report no verification result for that configuration. This is the expected wall of explicit-state model checking and it marks the boundary of what this approach certifies directly; verifying the pipeline for an unbounded number of endpoints and scans would require parameterized techniques, which we discuss as future work.

\section{Conclusion}
We modeled the endpoint decision pipeline of a distributed malware-detection system, abstracted from a production Bitdefender architecture, in Promela, and verified it with SPIN. The model captures what makes the system a \emph{malware-detection} system rather than a generic client--server one: a verdict is a consequential decision that the endpoint enforces locally, and the primary/legacy/local fallback chain must keep that decision safe when servers fail. Under explicit contracts for the full and reduced detection engines, we verified six LTL properties: the fallback machinery never causes a false positive or duplicate enforcement, weakens detection only through an explicit ordered descent, always reaches its single terminal commit point, and confines every false negative to the weakest fallback mode. A separate search establishes deadlock freedom, and reachability traps exercise the behaviors on which the conditional properties depend. We also report where exhaustive checking stops being tractable.

The verification targets the detection \emph{infrastructure}, the failure-handling logic that acts on verdicts, which is distinct from the malware-behavior analysis~\cite{song2014model} and the detection-rule verification~\cite{prelipcean2025bridging} that formal methods have previously brought to this domain.

Two directions stand out for future work. First, the state space of the concurrent configuration grows quickly, so certifying the pipeline for an unbounded number of endpoints and scans calls for parameterized techniques such as cutoff arguments or threshold automata~\cite{BaumeisterEJSV24}, rather than checking ever-larger finite instances. Second, our deadlines are untimed nondeterministic abandonment transitions; verifying quantitative timing properties (for instance, that a bounded deadline is always met) would call for a timed formalism and a tool such as UPPAAL~\cite{uppaal}.
\bibliographystyle{eptcs}
\bibliography{refs}

\begin{thebibliography}{10}
\providecommand{\bibitemdeclare}[2]{}
\providecommand{\surnamestart}{}
\providecommand{\surnameend}{}
\providecommand{\urlprefix}{Available at }
\providecommand{\url}[1]{\texttt{#1}}
\providecommand{\href}[2]{\texttt{#2}}
\providecommand{\urlalt}[2]{\href{#1}{#2}}
\providecommand{\doi}[1]{doi:\urlalt{https://doi.org/#1}{#1}}
\providecommand{\eprint}[1]{arXiv:\urlalt{https://arxiv.org/abs/#1}{#1}}
\providecommand{\bibinfo}[2]{#2}

\bibitemdeclare{inproceedings}{BaumeisterEJSV24}
\bibitem{BaumeisterEJSV24}
\bibinfo{author}{Tom \surnamestart Baumeister\surnameend},
  \bibinfo{author}{Paul \surnamestart Eichler\surnameend},
  \bibinfo{author}{Swen \surnamestart Jacobs\surnameend},
  \bibinfo{author}{Mouhammad \surnamestart Sakr\surnameend} \&
  \bibinfo{author}{Marcus \surnamestart V{\"{o}}lp\surnameend}
  (\bibinfo{year}{2025}): \emph{\bibinfo{title}{Parameterized Verification of
  Round-Based Distributed Algorithms via Extended Threshold Automata}}.
\newblock In \bibinfo{editor}{Andr{\'{e}} \surnamestart Platzer\surnameend},
  \bibinfo{editor}{Kristin~Yvonne \surnamestart Rozier\surnameend},
  \bibinfo{editor}{Matteo \surnamestart Pradella\surnameend} \&
  \bibinfo{editor}{Matteo \surnamestart Rossi\surnameend}, editors: {\slshape
  \bibinfo{booktitle}{Formal Methods - 26th International Symposium, {FM} 2024,
  Milan, Italy, September 9-13, 2024, Proceedings, Part {I}}}, {\slshape
  \bibinfo{series}{Lecture Notes in Computer Science}} \bibinfo{volume}{14933},
  \bibinfo{publisher}{Springer}, pp. \bibinfo{pages}{638--657},
  \doi{10.1007/978-3-031-71162-6_33}.

\bibitemdeclare{incollection}{clarke2012stateexplosion}
\bibitem{clarke2012stateexplosion}
\bibinfo{author}{Edmund~M. \surnamestart Clarke\surnameend},
  \bibinfo{author}{William \surnamestart Klieber\surnameend},
  \bibinfo{author}{Milo{\v s} \surnamestart Nov{\'a}{\v c}ek\surnameend} \&
  \bibinfo{author}{Paolo \surnamestart Zuliani\surnameend}
  (\bibinfo{year}{2012}): \emph{\bibinfo{title}{Model Checking and the State
  Explosion Problem}}.
\newblock In \bibinfo{editor}{Bertrand \surnamestart Meyer\surnameend} \&
  \bibinfo{editor}{Martin \surnamestart Nordio\surnameend}, editors: {\slshape
  \bibinfo{booktitle}{Tools for Practical Software Verification}}, {\slshape
  \bibinfo{series}{Lecture Notes in Computer Science}} \bibinfo{volume}{7682},
  \bibinfo{publisher}{Springer}, pp. \bibinfo{pages}{1--30},
  \doi{10.1007/978-3-642-35746-6_1}.

\bibitemdeclare{article}{abs-2505-11963}
\bibitem{abs-2505-11963}
\bibinfo{author}{Luca \surnamestart Collini\surnameend},
  \bibinfo{author}{Baleegh \surnamestart Ahmad\surnameend},
  \bibinfo{author}{Joey \surnamestart Ah{-}kiow\surnameend} \&
  \bibinfo{author}{Ramesh \surnamestart Karri\surnameend}
  (\bibinfo{year}{2025}): \emph{\bibinfo{title}{{MARVEL:} Multi-Agent {RTL}
  Vulnerability Extraction using Large Language Models}}.
\newblock {\slshape \bibinfo{journal}{CoRR}} \bibinfo{volume}{abs/2505.11963},
  \doi{10.48550/arXiv.2505.11963}.
\newblock \eprint{2505.11963v2}.
\newblock \bibinfo{note}{Preprint, version 2, 9 June 2025.
  \url{https://arxiv.org/abs/2505.11963v2}}.

\bibitemdeclare{inproceedings}{delzanno2014paxos}
\bibitem{delzanno2014paxos}
\bibinfo{author}{Giorgio \surnamestart Delzanno\surnameend},
  \bibinfo{author}{Michele \surnamestart Tatarek\surnameend} \&
  \bibinfo{author}{Riccardo \surnamestart Traverso\surnameend}
  (\bibinfo{year}{2014}): \emph{\bibinfo{title}{Model Checking {Paxos} in
  {Spin}}}.
\newblock In: {\slshape \bibinfo{booktitle}{Proceedings of the 5th
  International Symposium on Games, Automata, Logics and Formal Verification
  (GandALF)}}, {\slshape \bibinfo{series}{EPTCS}} \bibinfo{volume}{161}, pp.
  \bibinfo{pages}{131--146}, \doi{10.4204/EPTCS.161.13}.

\bibitemdeclare{inproceedings}{dwyer1999patterns}
\bibitem{dwyer1999patterns}
\bibinfo{author}{Matthew~B. \surnamestart Dwyer\surnameend},
  \bibinfo{author}{George~S. \surnamestart Avrunin\surnameend} \&
  \bibinfo{author}{James~C. \surnamestart Corbett\surnameend}
  (\bibinfo{year}{1999}): \emph{\bibinfo{title}{Patterns in Property
  Specifications for Finite-State Verification}}.
\newblock In: {\slshape \bibinfo{booktitle}{Proceedings of the 21st
  International Conference on Software Engineering (ICSE)}},
  \bibinfo{publisher}{ACM}, pp. \bibinfo{pages}{411--420},
  \doi{10.1145/302405.302672}.

\bibitemdeclare{incollection}{emersonLTL90}
\bibitem{emersonLTL90}
\bibinfo{author}{E.~Allen \surnamestart Emerson\surnameend}
  (\bibinfo{year}{1990}): \emph{\bibinfo{title}{Temporal and Modal Logic}}.
\newblock In \bibinfo{editor}{Jan \surnamestart van Leeuwen\surnameend},
  editor: {\slshape \bibinfo{booktitle}{Handbook of Theoretical Computer
  Science, Volume {B:} Formal Models and Semantics}},
  \bibinfo{publisher}{Elsevier and {MIT} Press}, pp.
  \bibinfo{pages}{995--1072}, \doi{10.1016/B978-0-444-88074-1.50021-4}.

\bibitemdeclare{article}{cadp}
\bibitem{cadp}
\bibinfo{author}{Hubert \surnamestart Garavel\surnameend},
  \bibinfo{author}{Fr\'ed\'eric \surnamestart Lang\surnameend},
  \bibinfo{author}{Radu \surnamestart Mateescu\surnameend} \&
  \bibinfo{author}{Wendelin \surnamestart Serwe\surnameend}
  (\bibinfo{year}{2013}): \emph{\bibinfo{title}{{CADP} 2011: A Toolbox for the
  Construction and Analysis of Distributed Processes}}.
\newblock {\slshape \bibinfo{journal}{International Journal on Software Tools
  for Technology Transfer}} \bibinfo{volume}{15}(\bibinfo{number}{2}), pp.
  \bibinfo{pages}{89--107}, \doi{10.1007/s10009-012-0244-z}.
\newblock \bibinfo{note}{\url{https://cadp.inria.fr/}}.

\bibitemdeclare{article}{holzmann2004spin}
\bibitem{holzmann2004spin}
\bibinfo{author}{G.J. \surnamestart Holzmann\surnameend}
  (\bibinfo{year}{1997}): \emph{\bibinfo{title}{The model checker SPIN}}.
\newblock {\slshape \bibinfo{journal}{IEEE Transactions on Software
  Engineering}} \bibinfo{volume}{23}(\bibinfo{number}{5}), pp.
  \bibinfo{pages}{279--295}, \doi{10.1109/32.588521}.

\bibitemdeclare{misc}{HowardKACC25}
\bibitem{HowardKACC25}
\bibinfo{author}{Heidi \surnamestart Howard\surnameend},
  \bibinfo{author}{Markus~A. \surnamestart Kuppe\surnameend},
  \bibinfo{author}{Edward \surnamestart Ashton\surnameend},
  \bibinfo{author}{Amaury \surnamestart Chamayou\surnameend} \&
  \bibinfo{author}{Natacha \surnamestart Crooks\surnameend}
  (\bibinfo{year}{2024}): \emph{\bibinfo{title}{Smart Casual Verification of
  the {Confidential Consortium Framework}}}, \doi{10.48550/arXiv.2406.17455}.
\newblock \eprint{2406.17455}.

\bibitemdeclare{inproceedings}{john2013fault}
\bibitem{john2013fault}
\bibinfo{author}{Annu \surnamestart John\surnameend}, \bibinfo{author}{Igor
  \surnamestart Konnov\surnameend}, \bibinfo{author}{Ulrich \surnamestart
  Schmid\surnameend}, \bibinfo{author}{Helmut \surnamestart Veith\surnameend}
  \& \bibinfo{author}{Josef \surnamestart Widder\surnameend}
  (\bibinfo{year}{2013}): \emph{\bibinfo{title}{Towards Modeling and Model
  Checking Fault-Tolerant Distributed Algorithms}}.
\newblock In: {\slshape \bibinfo{booktitle}{Model Checking Software (SPIN
  2013)}}, {\slshape \bibinfo{series}{LNCS}} \bibinfo{volume}{7976},
  \bibinfo{publisher}{Springer}, pp. \bibinfo{pages}{209--226},
  \doi{10.1007/978-3-642-39176-7_14}.

\bibitemdeclare{article}{kaur-2012}
\bibitem{kaur-2012}
\bibinfo{author}{Harpreet \surnamestart Kaur\surnameend} \&
  \bibinfo{author}{Amandeep \surnamestart Verma\surnameend}
  (\bibinfo{year}{2012}): \emph{\bibinfo{title}{{Formal modeling and
  verification of trusted OLSR protocol using I-SPIN Model Checker}}}.
\newblock {\slshape \bibinfo{journal}{IOSR Journal of Computer Engineering}}
  \bibinfo{volume}{4}(\bibinfo{number}{1}), pp. \bibinfo{pages}{01--05},
  \doi{10.9790/0661-0410105}.

\bibitemdeclare{inproceedings}{kupferman1999vacuity}
\bibitem{kupferman1999vacuity}
\bibinfo{author}{Orna \surnamestart Kupferman\surnameend} \&
  \bibinfo{author}{Moshe~Y. \surnamestart Vardi\surnameend}
  (\bibinfo{year}{1999}): \emph{\bibinfo{title}{Vacuity Detection in Temporal
  Model Checking}}.
\newblock In: {\slshape \bibinfo{booktitle}{Correct Hardware Design and
  Verification Methods (CHARME)}}, {\slshape \bibinfo{series}{LNCS}}
  \bibinfo{volume}{1703}, \bibinfo{publisher}{Springer}, pp.
  \bibinfo{pages}{82--98}, \doi{10.1007/3-540-48153-2_8}.

\bibitemdeclare{article}{uppaal}
\bibitem{uppaal}
\bibinfo{author}{Kim~G. \surnamestart Larsen\surnameend}, \bibinfo{author}{Paul
  \surnamestart Pettersson\surnameend} \& \bibinfo{author}{Wang \surnamestart
  Yi\surnameend} (\bibinfo{year}{1997}): \emph{\bibinfo{title}{{UPPAAL} in a
  Nutshell}}.
\newblock {\slshape \bibinfo{journal}{International Journal on Software Tools
  for Technology Transfer}} \bibinfo{volume}{1}(\bibinfo{number}{1--2}), pp.
  \bibinfo{pages}{134--152}, \doi{10.1007/s100090050010}.
\newblock \bibinfo{note}{\url{https://uppaal.org/}}.

\bibitemdeclare{inproceedings}{nakashiro2011translation}
\bibitem{nakashiro2011translation}
\bibinfo{author}{Ryosuke \surnamestart Nakashiro\surnameend},
  \bibinfo{author}{Yasutaka \surnamestart Kamei\surnameend},
  \bibinfo{author}{Naoyasu \surnamestart Ubayashi\surnameend},
  \bibinfo{author}{Shin \surnamestart Nakajima\surnameend} \&
  \bibinfo{author}{Akihito \surnamestart Iwai\surnameend}
  (\bibinfo{year}{2011}): \emph{\bibinfo{title}{Translation pattern of {BPEL}
  process into {Promela} code}}.
\newblock In: {\slshape \bibinfo{booktitle}{Proceedings of the Joint Conference
  of the 21st International Workshop on Software Measurement and the 6th
  International Conference on Software Process and Product Measurement
  (IWSM-MENSURA)}}, pp. \bibinfo{pages}{285--290},
  \doi{10.1109/IWSM-MENSURA.2011.42}.

\bibitemdeclare{inproceedings}{nardone2016modbus}
\bibitem{nardone2016modbus}
\bibinfo{author}{Roberto \surnamestart Nardone\surnameend},
  \bibinfo{author}{Ricardo~J. \surnamestart Rodr{\'i}guez\surnameend} \&
  \bibinfo{author}{Stefano \surnamestart Marrone\surnameend}
  (\bibinfo{year}{2016}): \emph{\bibinfo{title}{Formal Security Assessment of
  {Modbus} Protocol}}.
\newblock In: {\slshape \bibinfo{booktitle}{2016 11th International Conference
  for Internet Technology and Secured Transactions (ICITST)}},
  \bibinfo{publisher}{IEEE}, pp. \bibinfo{pages}{142--147},
  \doi{10.1109/ICITST.2016.7856685}.

\bibitemdeclare{article}{aws-correctness}
\bibitem{aws-correctness}
\bibinfo{author}{Chris \surnamestart Newcombe\surnameend}, \bibinfo{author}{Tim
  \surnamestart Rath\surnameend}, \bibinfo{author}{Fan \surnamestart
  Zhang\surnameend}, \bibinfo{author}{Bogdan \surnamestart
  Munteanu\surnameend}, \bibinfo{author}{Marc \surnamestart Brooker\surnameend}
  \& \bibinfo{author}{Michael \surnamestart Deardeuff\surnameend}
  (\bibinfo{year}{2015}): \emph{\bibinfo{title}{How {Amazon Web Services} Uses
  Formal Methods}}.
\newblock {\slshape \bibinfo{journal}{Communications of the ACM}}
  \bibinfo{volume}{58}(\bibinfo{number}{4}), pp. \bibinfo{pages}{66--73},
  \doi{10.1145/2699417}.

\bibitemdeclare{inproceedings}{arxiv-multigrained}
\bibitem{arxiv-multigrained}
\bibinfo{author}{Lingzhi \surnamestart Ouyang\surnameend},
  \bibinfo{author}{Xudong \surnamestart Sun\surnameend}, \bibinfo{author}{Ruize
  \surnamestart Tang\surnameend}, \bibinfo{author}{Yu~\surnamestart
  Huang\surnameend}, \bibinfo{author}{Madhav \surnamestart
  Jivrajani\surnameend}, \bibinfo{author}{Xiaoxing \surnamestart Ma\surnameend}
  \& \bibinfo{author}{Tianyin \surnamestart Xu\surnameend}
  (\bibinfo{year}{2025}): \emph{\bibinfo{title}{Multi-Grained Specifications
  for Distributed System Model Checking and Verification}}.
\newblock In: {\slshape \bibinfo{booktitle}{Proceedings of the Twentieth
  European Conference on Computer Systems, EuroSys 2025, Rotterdam, The
  Netherlands, 30 March 2025 - 3 April 2025}}, \bibinfo{publisher}{{ACM}}, pp.
  \bibinfo{pages}{379--395}, \doi{10.1145/3689031.3696069}.

\bibitemdeclare{inproceedings}{PraveenRD24}
\bibitem{PraveenRD24}
\bibinfo{author}{M.~\surnamestart Praveen\surnameend},
  \bibinfo{author}{Raghavendra \surnamestart Ramesh\surnameend} \&
  \bibinfo{author}{Isaac \surnamestart Doidge\surnameend}
  (\bibinfo{year}{2024}): \emph{\bibinfo{title}{Formally Verifying the Safety
  of {Pipelined Moonshot} Consensus Protocol}}.
\newblock In \bibinfo{editor}{Bruno \surnamestart Bernardo\surnameend} \&
  \bibinfo{editor}{Diego \surnamestart Marmsoler\surnameend}, editors:
  {\slshape \bibinfo{booktitle}{5th International Workshop on Formal Methods
  for Blockchains, {FMBC} 2024, April 7, 2024, Luxembourg City, Luxembourg}},
  {\slshape \bibinfo{series}{OASIcs}} \bibinfo{volume}{118},
  \bibinfo{publisher}{Schloss Dagstuhl - Leibniz-Zentrum f{\"{u}}r Informatik},
  pp. \bibinfo{pages}{3:1--3:16}, \doi{10.4230/OASIcs.FMBC.2024.3}.

\bibitemdeclare{inproceedings}{prelipcean2025bridging}
\bibitem{prelipcean2025bridging}
\bibinfo{author}{Dumitru-Bogdan \surnamestart Prelipcean\surnameend} \&
  \bibinfo{author}{C\u{a}t\u{a}lin \surnamestart Dima\surnameend}
  (\bibinfo{year}{2025}): \emph{\bibinfo{title}{Bridging Threat Models and
  Detections: Formal Verification via {CADP}}}.
\newblock In: {\slshape \bibinfo{booktitle}{Proceedings of the 9th Working
  Formal Methods Symposium (FROM 2025)}}, {\slshape \bibinfo{series}{EPTCS}}
  \bibinfo{volume}{427}, pp. \bibinfo{pages}{59--78},
  \doi{10.4204/EPTCS.427.5}.

\bibitemdeclare{inproceedings}{song2012efficient}
\bibitem{song2012efficient}
\bibinfo{author}{Fu~\surnamestart Song\surnameend} \& \bibinfo{author}{Tayssir
  \surnamestart Touili\surnameend} (\bibinfo{year}{2012}):
  \emph{\bibinfo{title}{Efficient Malware Detection Using Model-Checking}}.
\newblock In: {\slshape \bibinfo{booktitle}{FM 2012: Formal Methods}},
  {\slshape \bibinfo{series}{Lecture Notes in Computer Science}}
  \bibinfo{volume}{7436}, \bibinfo{publisher}{Springer}, pp.
  \bibinfo{pages}{418--433}, \doi{10.1007/978-3-642-32759-9_34}.

\bibitemdeclare{inproceedings}{song2014model}
\bibitem{song2014model}
\bibinfo{author}{Fu~\surnamestart Song\surnameend} \& \bibinfo{author}{Tayssir
  \surnamestart Touili\surnameend} (\bibinfo{year}{2014}):
  \emph{\bibinfo{title}{Model-Checking for {Android} Malware Detection}}.
\newblock In \bibinfo{editor}{Jacques \surnamestart Garrigue\surnameend},
  editor: {\slshape \bibinfo{booktitle}{Programming Languages and Systems: 12th
  Asian Symposium, {APLAS} 2014}}, {\slshape \bibinfo{series}{Lecture Notes in
  Computer Science}} \bibinfo{volume}{8858}, \bibinfo{publisher}{Springer
  International Publishing}, \bibinfo{address}{Cham}, pp.
  \bibinfo{pages}{216--235}, \doi{10.1007/978-3-319-12736-1_12}.

\bibitemdeclare{article}{xiao2022composition}
\bibitem{xiao2022composition}
\bibinfo{author}{Meihua \surnamestart Xiao\surnameend}, \bibinfo{author}{Hanyu
  \surnamestart Zhao\surnameend}, \bibinfo{author}{Ke~\surnamestart
  Yang\surnameend}, \bibinfo{author}{Ri~\surnamestart Ouyang\surnameend} \&
  \bibinfo{author}{Weiwei \surnamestart Song\surnameend}
  (\bibinfo{year}{2022}): \emph{\bibinfo{title}{A formal analysis method for
  composition protocol based on model checking}}.
\newblock {\slshape \bibinfo{journal}{Scientific Reports}}
  \bibinfo{volume}{12}, p. \bibinfo{pages}{8493},
  \doi{10.1038/s41598-022-12448-2}.

\bibitemdeclare{article}{yang2022formal}
\bibitem{yang2022formal}
\bibinfo{author}{Zhe \surnamestart Yang\surnameend}, \bibinfo{author}{Meiyi
  \surnamestart Dai\surnameend} \& \bibinfo{author}{Jian \surnamestart
  Guo\surnameend} (\bibinfo{year}{2022}): \emph{\bibinfo{title}{Formal Modeling
  and Verification of Smart Contracts with {Spin}}}.
\newblock {\slshape \bibinfo{journal}{Electronics}}
  \bibinfo{volume}{11}(\bibinfo{number}{19}), p. \bibinfo{pages}{3091},
  \doi{10.3390/electronics11193091}.

\bibitemdeclare{misc}{zhang2025languageagnosticlogicalrelationmessagepassing}
\bibitem{zhang2025languageagnosticlogicalrelationmessagepassing}
\bibinfo{author}{Tesla \surnamestart Zhang\surnameend}, \bibinfo{author}{Sonya
  \surnamestart Simkin\surnameend}, \bibinfo{author}{Rui \surnamestart
  Li\surnameend}, \bibinfo{author}{Yue \surnamestart Yao\surnameend} \&
  \bibinfo{author}{Stephanie \surnamestart Balzer\surnameend}
  (\bibinfo{year}{2025}): \emph{\bibinfo{title}{A Language-Agnostic Logical
  Relation for Message-Passing Protocols}}, \doi{10.48550/arXiv.2506.10026}.
\newblock \eprint{2506.10026}.
\newblock \bibinfo{note}{Preprint, arXiv:2506.10026v1.
  \url{https://arxiv.org/abs/2506.10026v1}}.

\end{thebibliography}

\end{document}